%% file: graph-based-gmt.tex
\documentclass[12pt]{easychair}

\usepackage[colorlinks,linkcolor=blue,urlcolor=blue,citecolor=blue]{hyperref}
\usepackage{color}
\usepackage{enumerate}
\usepackage[subfigure]{tocloft}
\usepackage{listings}
\usepackage{subfigure}
\usepackage{fancyvrb}

\usepackage{paralist}

\lstdefinestyle{codeStyle}{
    captionpos=b,%
    showstringspaces=false,%
    showspaces=false,
    frame=single,  
    extendedchars=true,%
    basicstyle=\scriptsize\tt, 
    linewidth=1\linewidth,%
    language=Java,%
    breaklines=true,
    float=phtb,  
}

\input{commands}

\newif\ifarxiv
\arxivtrue

\newif\iftoc
\toctrue

\newcommand
	{\deleteifsmaller}%
	[1]
	{\ifarxiv#1\else\fi}

\begin{document}





\ifarxiv

\title{An Interactive Graph-Based Automation Assistant:
A Case Study to Manage the GIPSY's Distributed Multi-tier Run-Time System}
\titlerunning{Graph-Based Tool To Manage GIPSY Networks}

\author
{
	Sleiman Rabah, Serguei A. Mokhov and Joey Paquet \\
	Computer Science and Software Engineering\\
	Concordia University, Montreal, QC, Canada\\
	\url{{s_rabah,mokhov,paquet}@encs.concordia.ca}
}
\authorrunning{Rabah, Mokhov, Paquet}

\else

\title{An Interactive Graph-Based Automation Assistant:
A Case Study to Manage the GIPSY's Distributed Multi-tier Run-Time System}
\short{Graph-Based Tool To Manage GIPSY Networks}

\author
{
	Sleiman Rabah\autref{1},
	Serguei A. Mokhov\autref{1},
	Joey Paquet\autref{1}
}

\institute
{
	\autlabel{1}Computer Science and Software Engineering\\
	Concordia University, Montreal, QC, Canada\\
	\email{{s\_rabah,mokhov,paquet}@encs.concordia.ca}
}

\fi

\ifarxiv\else
\fi

\maketitle

%
\begin{abstract}
The {\gipsy} system provides a framework for a distributed multi-tier demand-driven 
evaluation of heterogeneous programs, in which certain tiers can generate 
demands, while others can respond to demands to work on them. They are connected through a 
virtual network that can be flexibly reconfigured at run-time. Although the 
demand generator components were originally designed specifically for the eductive (demand-driven)
evaluation of Lucid intensional programs, the {\gipsy}'s run-time's flexible framework 
design enables it to perform the execution of various kinds of programs that 
can be evaluated using the demand-driven computational model. Management
of the GISPY networks has become a tedious (although scripted) task that
took manual command-line console to do, which does not scale for
large experiments.
Therefore a new component has been designed and developed to allow users to represent, 
visualize, and interactively create, configure and seamlessly manage such a network
as a graph.
Consequently, this work presents a Graphical GMT Manager, an interactive graph-based assistant component
for the GIPSY network creation and configuration management. Besides allowing the management
of the nodes and tiers (mapped to hosts where store, workers, and generators reside), it lets the user
to visually control the network parameters and the interconnection between
computational nodes at run-time. In this paper we motivate and present the 
key features of this newly implemented graph-based component. We give the graph representation
details, mapping of the graph nodes to tiers, tier groups, and specific commands. We provide
the requirements and design specification of the tool and its implementation.
Then we detail and discuss some experimental results.
\ifarxiv
\fi
\end{abstract}

\iftoc
	\clearpage
	\tableofcontents
	\listoffigures
	\clearpage
\fi

\section{Introduction}
\label{sect:introduction}	

The GIPSY (General Intensional Programming System) project is an ongoing 
research project developed at Concordia University. Its initial goal was to 
investigate on a general solution for the evaluation of programs written in 
the {\lucid} intensional programming family of languages using a distributed 
demand-driven evaluation model. In order to meet the flexibility goals 
of the project, the system has been designed using a framework approach 
integrating a Lucid compiler framework, as well as a demand-driven run-time 
system framework.

In its eductive model of execution, the system assumes the presence of demand 
generators, as well as a demand workers. Each demand generated is paired 
along with the context in which it is made and is uniquely identified. The 
demands are migrated using a communication node that enables the connection 
between demand generators and demand workers. Through the communication node, 
any demand worker can pick-up demands, compute its resulting value, and send 
it back to the communication node to be picked up by the generator. 

Notably, the framework has demonstrated its flexibility by having the run-time
system put to use in the demand-driven distributed evaluation of 
programs not involving the Lucid language. The work presented here goes in 
this direction and makes abstraction of the intensional programming aspect of 
the project and concentrates on the demand-driven evaluation of heterogeneous 
programs. We concentrate on showing how a virtual network of demand-driven 
computational nodes can be represented graphically at run-time, enabling the 
user to map the demand-driven computation nodes over an underlying physical 
network of computers, and to control their execution and connectivity at run-time.

In the current implementation, the node's connectivity is expressed in a set 
of configuration files. Upon starting, a node reads its configuration file 
and establishes its own connection according to the information contained in 
the configuration file. The configuration can be changed at any time, so that 
a node can reconfigure its connectivity at run-time. 

A manager node, which acts as a supernode, has been implemented to manage a {\gipsynetwork}.
It enables new node(s) to automatically establish connection with 
it and receive commands from it. Therefore, enabling the manager node to remotely change 
the configuration of any registered node. The virtual network is thus 
constructed from the interconnection of the generators, workers, 
communication, and manager nodes, the later being able to establish the 
connectivity between the three first ones. All communication between nodes, 
including commands exchanged for configuration changes, are using the same 
demand-driven communication mode.    

The rest of this paper is organized as follows: \xs{sect:background} gives on overview 
of the GIPSY Framework and its multi-tier architecture, the GIPSY run-time 
system and finally discusses the related work. \xs{sect:objectives} summarizes the 
objectives 
of this work. \xs{sect:methodology} presents how currently the GIPSY run-time system is 
being managed, discusses the design and implementation of the proposed 
solution and evaluates the results of some conducted experiments. 
Then, \xs{sect:conclusion} concludes the paper and points out new research direction 
planned as future work.

\section{Background}
\label{sect:background}

\subsection{GIPSY Framework}
\label{sect:bg-gipsy-framework}

The GIPSY run-time system is a distributed multi-tier and demand-driven 
framework. It mainly consists of a set of loosely coupled software components 
enabling the evaluation of programs in a distributed demand-driven manner. 
The run-time system is composed of the following basic entities ~\cite{gipsy-multi-tier-secasa09}:
(a) \textbf{A {\gipsytier}} is an abstract and generic entity. Each tier instance is a separate thread (one or more) that runs within a registered 
process, namely ({\gipsynode}), and represents a computational unit that contribute to the distributed computation. 
Tiers cooperate in a demand-driven mode of computation;
(b) \textbf{A {\gipsynode}} is a registered process that hosts 
one or more GIPSY tier instances belonging to different {\gipsyinstance}(s). Node registration is done through a manager 
tier called the GIPSY Manager Tier ({\gmt}). More specifically, a node is a 
computer running a \textit{GIPSYNode} process;
(c) \textbf{A {\gipsyinstance}}
is a group of tier instances collaborating together to achieve program 
execution. It can be considered as a set of interconnected GIPSY tier 
instances hosted/deployed on one or more GIPSY nodes executing GIPSY programs 
by sharing their respective resources. A GIPSY instance can be executed 
across different GIPSY nodes. Moreover, as shown in~\xf{fig:gipsy-nodes-network},
a {\gipsynetwork} is designed as an overlay network where network nodes, {\gipsytier}s, are 
organized in a cluster called {\gipsyinstance}. A {\gipsytier} can be 
seen as a virtual network node and hosted on a {\gipsynode}. In such a 
network, the mapping between a {\gipsynode} and a physical node is made upon 
starting and registering the node through the {\gmt}. 

\subsubsection{Multi-Tier Architecture}
\label{sect:bg-multi-tier-architecture}

In~\cite{gipsy-multi-tier-secasa09}, a distributed multi-tier architecture has 
been defined and adopted in the implementation of GIPSY run-time system. The architecture 
inherits some of the peer-to-peer network architecture principles, e.g (1) no 
single-point of failure: any tier or node can fail without fatally affecting 
the system; (2) nodes and tiers can seamlessly join/leave the network by 
adding/removing them on the fly as computation is happening; (3) demands are 
propagated without knowing where they will be processed or stored; (4) available
nodes and tiers can be affected at run-time to the execution of any 
{\gipsy} program while other nodes and tiers could be computing demands for 
different programs. The multi-tier architecture is composed of four distinct 
tiers: (a) a Demand Store Tier ({\dst}) that acts as a middleware between tiers in order to migrate demands, 
provides persistent storage of demands and their resulting values (demands 
caching), and exposes Transport Agents (TAs) used by other tiers to connect to 
the {\dst}; (b) a Demand Generator Tier ({\dgt}) that generates demands according to the declarations and 
resources stored in the GEER generated for the program being evaluated. The
{\dgt} maintains a local demand processing dictionary pool that contain the 
definitions required to formulate demands; (c) a Demand Worker Tier ({\dwt}) which processes demands by 
executing method defined in such a dictionary. The {\dwt} connects to the
{\dst}, retrieves pending demands and returns back the computed demands to the 
{\dst}; (d) a General Manager Tier ({\gmt}) (see 
\xf{fig:gipsy-nodes-network}), as its name implies, locally and remotely controls and 
monitors other tiers ({\dgt}, {\dwt} and {\dst}) by exchanging system 
demands. Furthermore, the {\gmt} can register new nodes, move tier instances 
from one node to another, or allocate/deallocate tier instance from/on a 
registered node. 
 
\begin{figure}[htpb]%
	\centering
	\includegraphics[width=.7\columnwidth]{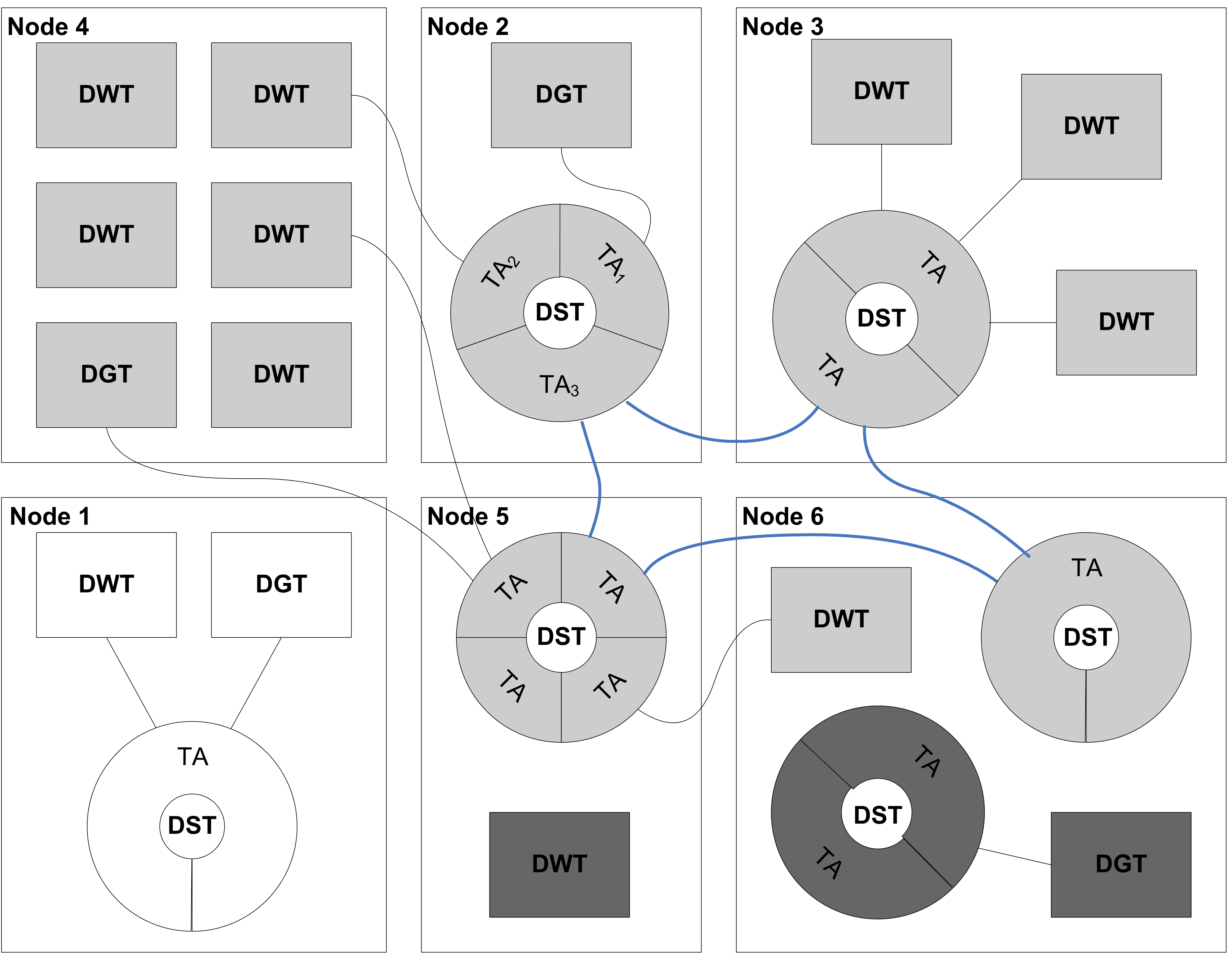}%
	\caption{Example of a GIPSY Nodes Network \protect~\cite{gipsy-multi-tier-secasa09}}%
	\label{fig:gipsy-nodes-network}%
\end{figure}

\subsubsection{Demand Types and States}
\label{sect:bg-demand-types-states}

A demand is a run-time request asking for a value 
of certain identifier defined in a GIPSY program.
Demands are migrated to other tiers using the {\dst}.
There are three types of demands:
(a) intensional demands, which are generated for the evaluation of a GIPSY 
program by a 
generator. GIPSY programs are written in a declarative style, where an 
identifier is defined as an expression using other identifiers defined in a 
multidimensional context space~\cite{gipsy-multi-tier-secasa09}. All demands 
for GIPSY identifiers contain the context in which they are made, and their 
evaluation depend on this context. The GIPSY program also uses an algebra of 
procedures that can be called during evaluation, which are called at run-time 
and become procedural demands;
(b) procedural demands which are generated by the {\dgt} when it encounters
a procedural function call during the GIPSY program 
evaluation. Procedural demands are processed by the {\dwt};
(c) system demands, in turn, are issued by the {\gmt} for run-time management 
purposes and include 
demands for monitoring and controlling tiers at run-time. It is worth 
mentioning that system demands are 
requests for managerial tasks e.g. demand for node registration, tier 
allocation and deallocation. In contrast, intensional demands and procedural 
demands are computational demands, i.e. demands that are generated during the 
evaluation process of a specific GIPSY program. 
	
In the {\gipsy} environment, each demand has a state and demand states are 
used to manage and propagate demands. The state transitions are managed by 
the demand store tier {\dst}, who is responsible for demand migration between 
generator and worker tiers. We distinguish three possible states as follows:
	(a) \textbf{Pending:} a pending demand is a demand that has been issued by 
a tier to the demand 
store and not yet picked by another tier for further processing. Pending 
demands are sent to generators and workers when they notify their 
availability for processing \cite{gipsy-multi-tier-secasa09}; 
	(b) \textbf{Processing:} a processing demand is a demand that has been 
grabbed by a tier from the 
demand store and its evaluation still being processed. This state is assigned 
by the demand store in order to make sure that the same demand is grabbed by 
only one tier for processing. When a tier goes out of service, all its 
associated processing demands are put back to the pending state by the demand 
store, ensuring a fail-safe behavior \cite{gipsy-multi-tier-secasa09};  
	(c) \textbf{Computed:} indicates that a demand has been computed. 
When a tier grabs a demand and is finished computing its corresponding 
value, it sends back the 
result to the demand store, that stores it in place of the initial demand and 
marks it as computed. Any further demands generation with the same context
will result in the store to directly respond with the 
resulting value, thus saving computation time \cite{gipsy-multi-tier-secasa09}.

\subsection{Graphical GMT Tool Support for the GIPSY Run-time System}
\label{sect:GIPSYRunTimeSystem}

Here we provide some details how the graph-based tool we developed assists with the
automation of the startup sequence and management tasks of the system.

\paragraph{Configuration.}
The tier instantiation process has a flexible design and has been implemented using Java Reflection~\cite{java-reflection}
and the Factory design pattern~\cite{head-first-patterns-2004}. It uses a configuration-based system
to instantiate tiers on the fly. Generic configuration instances are stored in files and their settings
can be easily updated and tailored to a specific tier's implementation requirements. Configurations contain
a set of key-value pairs where the key denotes the name of the configuration property while the values could be anything
from a service name, port number, IP address, etc. Such a configuration system eliminates the need of writing
or adapting source code to reflect a specific tier configuration. The properties stored in the \api{Configuration}
object determine the tier class to instantiate and consist of different settings interpreted by the tier
implementation class. Upon receiving a tier instantiation request, the \api{TierFactory} inspects the
configuration instance to determine which tier implementation class to instantiate using
Java Reflection~\cite{java-reflection}.

\paragraph{Bootstrapping.}

as mentioned earlier, a {\gipsynetwork} consists of a set of interconnected GIPSY 
nodes each hosting {\gipsytier}s mapped to physical machines where the GIPSY 
run-time is deployed. Such a network is managed by a {\gipsy} manager tier 
({\gmt}) that enables nodes registration to the network and tier allocation 
on the registered nodes. The bootstrap process of the GIPSY manager tier 
starts a registration demand store that is used solely for the exchange of 
system demands with the nodes and tiers allocated in the GIPSY network 
managed by this manager tier. Thus, system demands and computational demands 
are exchanged using different communication channels. Any computer deploying 
the GIPSY node run-time system can send a registration request to the manager 
tier, enabling this manager tier to remotely connect and control the 
allocation of various tiers on the registered nodes. After tiers are 
allocated to the registered nodes, the manager tier can connect the different 
tiers together, and eventually instruct a generator to start the demand-driven
evaluation of a GIPSY program. Even after execution is started, the 
manager tier can accept new nodes registrations, or allocate/deallocate new 
tiers on any registered node that it manages and make newly allocated tiers 
to contribute to a program's evaluation on the
fly~\cite{unifying-refactoring-jini-jms-dms,ji-yi-mcthesis-2011}.

\paragraph{GIPSY Node Registration.}

When a node wants to join the network, first, a {\gipsynode} issues a request 
to the {\gmt} expressed as \api{NodeRegistration} system demand having \textit
{pending} as state. Upon receiving a \api{NodeRegistration} demand, the {\gmt}
 assigns a {\dst} to the {\gipsynode} who issued the request. Afterward, the {
\gmt} saves the node registration information in a \api{GMTInfoKeeper} object 
and sends back a \api{RegistrationResult} demand having \textit{computed} as 
state and containing the DST information and the assigned node ID. Finally, 
the {\gipsynode} processes the result and uses the information contained in 
the demand to establish a connection to the assigned {\dst}. Establishing 
such connection creates a communication channel for further exchange of 
system demands.

\paragraph{Tier Allocation.}

Tiers are allocated inside a previously registered {\gipsynode}. The process of tier
allocation is performed through the operation of the {\gmt} using a pair of system demands:
\api{TierAllocationRequest} and \api{TierAllocationResult}. Both demands share the
same demand signature but have different states: \textit{pending} and \textit{computed}
respectively. The following information needs to be specified in the \api{TierAllocationRequest}
demand: the node identifier of the {\gipsynode} where the tiers to be allocated, the type of
the tier and how many tier instances are to be allocated. When the allocation process
is completed a \api{TierAllocationResult} demand is triggered and contains a set of tier
registrations. Each tier registration contains information such as the tier identifier,
which is internally assigned by the {\gipsynode}.

\paragraph{Tier Deallocation.}

Tier deallocation consists of removing previously allocated tiers. Similarly to the case of the tier
allocation process, two system demands \\ \api{TierDeallocationRequest}  and \api{TierDeallocationResult}
are issued by the {\gmt} to deallocate tiers upon user's request. The type and the ID of the tier
to be deallocated are embedded in a \api{TierDeallocationRequest} and sent to the \api{GIPSYNode}
process to deallocate the tier specified.  

\subsection{Related Work}

Related work by several researchers on visualization of load balancing, configuration,
formal systems for diagrammatic modeling and visual languages
and the corresponding graph systems are presented in
\cite{%
sim-viz-resource-alloc-control,%
visual-config-representation,%
logical-reasoning-with-diagrams,%
graph-transform-visual-languages,%
diagramatic-formal-system-euclidean}.
They all define key concepts that
are relevant to our visualization mechanisms within {\gipsy}
and its corresponding General Manager Tier~\cite{ji-yi-mcthesis-2011}.

\section{Objectives}
\label{sect:objectives}

The {\gipsy} framework has been designed in a modular manner but has a lot of configurable components;
hence, the need of  an automation solution for configuring and managing GIPSY deployment 
components is crucial. Moreover, prior to this work, the run-time system was 
managed using primarily a command-line interface. 
The project should provide an integrated tool that allows the user to:
%
\begin{inparaitem}[]
\item create a {\gipsynetwork} and configure its components (GIPSY instances, tiers and nodes);
\item save/load a GIPSY network configuration;
\item start/stop GIPSY nodes and register them with the {\gmt}, and allocate/deallocate GIPSY tiers;
\item dynamically visualize GIPSY nodes and tiers and inspect/change their properties at run-time;
\item change tiers connectivity at run-time;
\item increase the usability of {\gipsy} run-time system as a whole;
\item provide means and semantics for scheduling, validation, and visual mapping to {\lucid} programs.
\end{inparaitem}
%
The {\gmt} is the central element of our system from the user's perspective.
It enables to handle the managerial tasks related to the configuration and 
functioning of a GIPSY network. The proposed solution should be transparent 
and efficient enough in order to enhance the system usability, flexibility, 
and end-users experience, while maintaining the structure for run-time
analysis and scheduling.

\section{Solution}
\label{sect:methodology}

\subsection{Overview}

The solution presented in this paper is a graph-based graphical user interface that 
provides a set of user interfaces enabling the users to directly interact 
with the distributed {\gipsy} run-time system. The main objectives (cf. \xs{sect:objectives}) of this work consist of 
increasing the usability of the run-time system and enabling the user to have 
full control over the {\gipsynetwork} with a minimum of detailed manual intervention.
It should be noted that, prior to this work, all the managerial and 
configuration tasks needed to bootstrap a {\gipsynetwork} required the user 
to manually execute shear number of commands and scripts. In this work, we 
designed the graphical {\gmt} component that aims at allowing the user to 
manage and operate the entire {\gipsynetwork} seamlessly by translating 
simple graphical user interactions into complex message passing between the 
underlying deployment components. Our solution enables the user to easily 
create, configure, and control a {\gipsynetwork} through a graph-based 
interface. GIPSY tiers are illustrated as connected graph nodes. Tiers'
properties are read from files and stored as \api{Configuration} objects 
embedded in the graph nodes. We use graph element shapes to differentiate GIPSY instances 
and colors to differentiate GIPSY nodes. When the user adds a new tier to the
network graph, the color assigned to the tier is associated to the node the tier is assigned to.

According to the {\gipsy} multi-tier architecture, the {\dwt}, {\dgt} and
{\dst} expose software interfaces to be used for their mutual interactions. 
Since the {\gmt} plays a key role in the GIPSY network management, it provides a 
handy mechanism for starting and stopping nodes, and allocating and 
deallocating tiers. In the current run-time implementation, the interaction 
with the {\gmt} is command-based and is done through a command-line console UI,
with which the user manually bootstraps and controls the nodes 
and tiers by entering commands the corresponding. Additionally, a set of configuration files 
with the appropriate settings and properties for each tier type are needed. 
Before performing any node or tier startup or registration, we assume that a 
set of configuration files with appropriate settings and properties for nodes 
each tier type have been created. Typically, in order to start a network, the following 
sequence of steps should normally be performed:
%
\begin{inparaenum}
\item At first, a {\gipsynode} process should be created; that prompts the 
user to start a {\gmt} tier. When a {\gmt} is started, the {\gipsynode} is 
automatically registered and a registration {\dst} is allocated \cite{ji-yi-mcthesis-2011}. The 
registration {\dst} enables the {\gmt} to receive system demands for further 
node and tier allocations. This is the initial bootstrapping 
process that enables all further operations on a GIPSY network. 
\item At any time the user can expand the network by adding an additional 
node locally on the computer where the {\gmt} is executing (and tiers on a remote 
computer). Upon successful node creation, the user is prompted to register the 
node to an existing {\gmt} using the \api{register} command.
\item Then, on any registered node, {\dst}s are started to allow the 
propagation of demands between generators and workers, {\dgt}s are registered 
to generate demands, and {\dwt}s are registered to process the generated 
demands. 
\end{inparaenum}

Based on this discussion,
the following is a typical list of example commands that are used to interact with 
the run-time system in order to setup a manual {\gipsy} network:

\begin{enumerate}

\item {\texttt{start GMT GMTConfigFile.config}}\\*
This command starts the 
bootstrap process explained above, where GMT is the type of the tier and
\api{GMTConfigFile.config} is the configuration file that contains the settings 
and properties needed to instantiate a {\gmt} tier instance.

\item {\texttt{allocate NodeID TierType TierTypeConfigFile DSTIndex HowMany}}\\*
This command allocates a {\dgt} or a {\dwt}. \textit{NodeID} is 
the numeric ID of the node where the tier should be allocated, \textit{TierType} is 
the type of the tier ([{\dgt}, {\dwt}]), \textit{DSTIndex} is the index of 
the {\dst}, to which the tier in question should connect to and the
\textit{TierTypeConfigFile} is the tier-specific configuration file to use.

\item {\texttt{allocate NodeID DST DSTConfigFile.config HowMany}}\\*
This command sends a request to the {\gmt} with the node ID where a DST 
instance will be allocated, how many {\dst} instances are needed, and a {\dst}
configuration file name.

\item \texttt{deallocate NodeID TierType TierID1 TierID2 TierID\textit{n}}\\*
This command issues a demand to be processed by the {\gmt} to 
deallocate tiers. \textit{TierType} is the type of the \textit{tierID[1..n]} 
is tier instances IDs to deallocate in a node specified by its ID.
\end{enumerate}

The GIPSY network configuration process requires the user not only to know 
all the commands and their exact syntax, but requires to keep track of the 
IDs of the nodes and tiers.
It also requires the user to manually edit the related configuration files.
The configuration files contain many 
configuration elements that are not of importance in the node/tier management 
process, thus leading to confusion and possible mistakes. Our newly designed 
graphical GMT assistant rather allows the user to abstractly manipulate icons and use menu 
options to effectuate these operations. These GUI operations initiated by the 
user are then translated by the graphical GMT into the commands similar to the presented earlier.
As for the changes to the configuration files, the user is presented through 
the GUI with only the configuration elements that are relevant in the context 
of use, thus reducing the information load on the user and reducing the 
possibility of configuration mistakes. \xl{list:gispy-DGT-config} shows 
the content of a configuration file for a {\dgt} \cite{ji-yi-mcthesis-2011}. It provides configuration 
information such as to which class implementation to instantiate, the number of 
instances to be created, the mode of communication to use, and a maximum number of
demands that can be generated. These configuration parameters are read during 
startup and will determine the behavior of the generator. 

\begin{lstlisting}[
    label={list:gispy-DGT-config},
    caption={A Sample of DGT Configuration File},
    style=codeStyle
    ]
# Which implementation of the DGT class to instantiate.
gipsy.GEE.multitier.wrapper.impl=gipsy.tests.GEE.simulator.DGTSimulator
gipsy.GEE.multitier.DGT.DemandDispatcher.impl=gipsy.GEE.IDP.DemandGenerator.jini.rmi.JiniDemandDispatcher
# 0 Concurrent asynchronously
# 1 User-controlled asynchronously
# 2 Response time tester: synchronously
# 3 Space-scalability tester.
gipsy.tests.GEE.simulator.mode=2
gipsy.tests.GEE.simulator.tester.parameter=1
# Number of instances to be created.
gipsy.tests.GEE.simulator.tester.number=2
# Number of maximum demands.
gipsy.tests.GEE.simulator.demand.payload=32
\end{lstlisting}

\subsection{Design and Implementation}

Our implementation relies on the graph-based visualization to illustrate a
{\gipsynetwork}. We represent a {\gipsytier}s network as interconnected graph nodes
where each such node contains data/properties used in tier-to-tier 
communication configuration. Such properties are assigned and configured by 
the user when creating a {\gipsynetwork}. The GMT GUI was implemented using 
the Java JFC/Swing library. The {\gipsy}'s \api{Configuration} class is used 
to store different components' configuration. We have selected the Java 
Universal Network/Graph (JUNG) library to implement the visualization
of the management aspect of GIPSY nodes~\cite{jung-url,o-madadhain-fisher-2003}. 
JUNG is an open-source library for modeling data that can be represented as a 
graph or a network. JUNG provides many visualization features that can be 
changed at runtime such as node color, shape, and size. Thus, graph nodes can 
be grouped together, which enables us to differentiate the nodes by their tier
type ({\dst}, {\gmt}, or {\dwt}). Through JUNG, GIPSY nodes are configured 
while creating a connected graph of nodes and to visualize and manage their 
activities to alleviate manual complexity of such operations. The GMT GUI 
addresses the need of the automation of the managerial tasks of the GIPSY
run-time system and the configuration of resources.

The implemented features are:
%
\begin{inparaenum}
  \item create a new {\gipsynetwork} as a graph; 
  \item save/load a pre-configured {\gipsynetwork} to/from files;
  \item start, register and stop {\gipsynode}s by maintaining a color-differentiated
list of nodes with their related commands and configurations 
available in a context menu; 
  \item allocate and deallocate {\dst}s, {\dgt}s and {\dwt}s by manipulating 
icons and context menus;
  \item start/stop the demand-driven evaluation process on a {\dgt} trough a 
contextual menu accessed on its icon. 
\end{inparaenum}

The process of node registration and tier allocation has been embedded into 
our tool, and only the most relevant configuration information is shown to the 
user. Graphical objects representing GIPSY nodes encapsulate their related 
commands and hold the necessary properties for user inspection or change at 
setup or run-time. As for {\gipsytier} graphical objects, in addition to 
allocate and deallocate commands, these objects provide a drag-and-drop 
mechanism used to change the connectivity between tiers on the fly at run-time
(see \xf{fig:gmt-operator-view}). When a new tier or a {\gipsynode} is 
added to the network, it is automatically pre-configured and associated with a 
configuration file with the properties entered by the user
(see \xf{fig:network-graph-editor}). The GMT GUI is arranged in a tabbed form and 
provides two main distinct editor and operator views.

The network graph editor and resource configuration allow 
to create a {\gipsynetwork} or load an existing one. As shown in
\xf{fig:network-graph-editor}, {\gipsy} instances and nodes are arranged in two 
lists while {\gipsytier}s in a graph illustrating interconnected tiers. 
Instances, nodes and tiers can be easily added and configured separately. The 
configuration process is completely automated using dialog boxes allowing the 
user to fill in the configuration properties of each entity. All data entered 
is validated allowing only valid values to be accepted. In this 
editor, two {\gipsytier}s could be connected together and their configuration 
commands automatically generated by drawing a line to connect two graph nodes.

The {\gmt} operator lists context-menu-enabled {\gipsynode}s allowing
the user to start or stop {\gipsynode}s and register 
them with the {\gmt} by simple mouse clicks. As illustrated
in~\xf{fig:gmt-operator-view}, {\gipsytier}s are shown as connected graph nodes. Tiers 
belonging to the same {\gipsyinstance} are assigned the same shape. The 
tier's color determines on which {\gipsynode} a given tier is hosted on. 
Moreover, inspection and visualization of any element's properties is 
possible at run-time. This enables the user, for instance, to know which
{\gipsytier} is residing on which {\gipsynode}. The run-time system activities 
such as the output of {\gmt}, {\gipsynode}s and {\gipsytier}s, errors, and 
log messages are displayed in a separate distinct views. This provides better 
failure traceability and errors troubleshooting while, at the same time, 
providing useful information related to the overall computation process.

%
%

\begin{figure}
\centering
\mbox{
 \subfigure[GMT Operator View]{
    \includegraphics[width=3in]{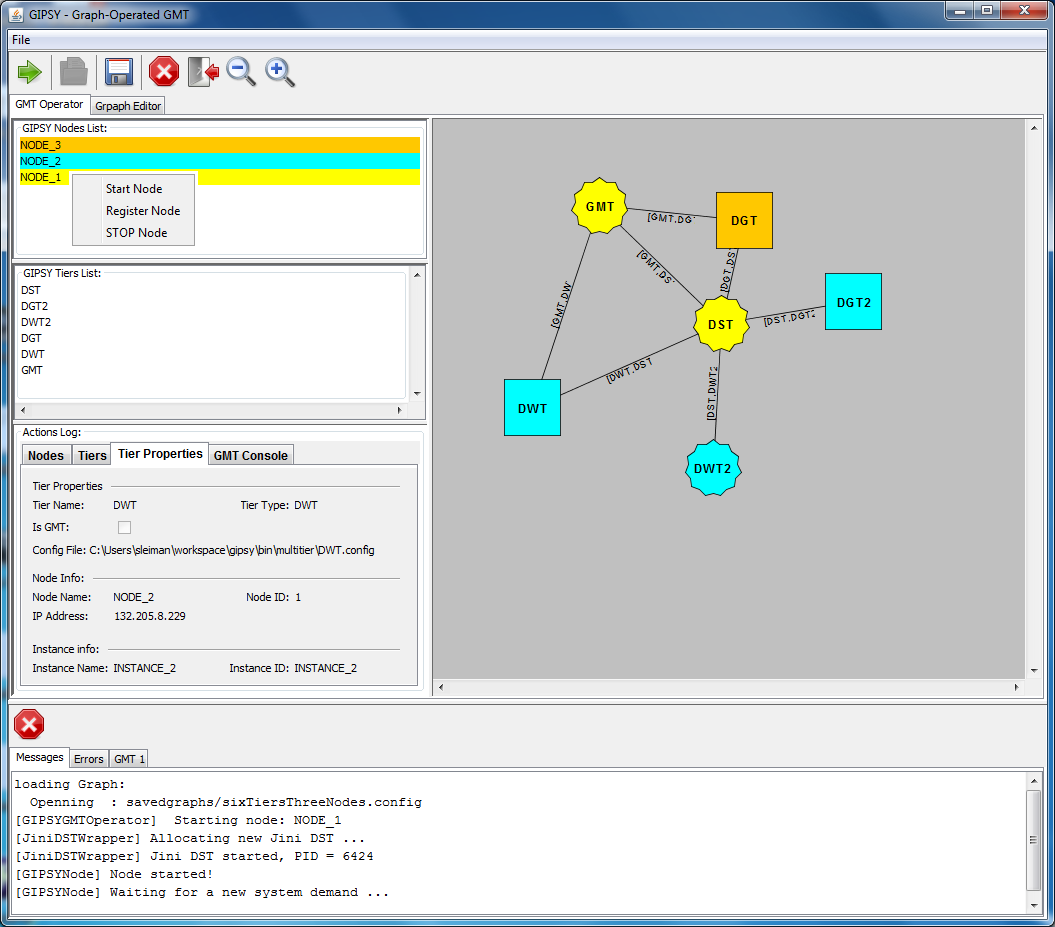}
		\label{fig:gmt-operator-view}
		}
    \quad
    \subfigure[Network Graph Editor View]{
		   \includegraphics[width=3in]{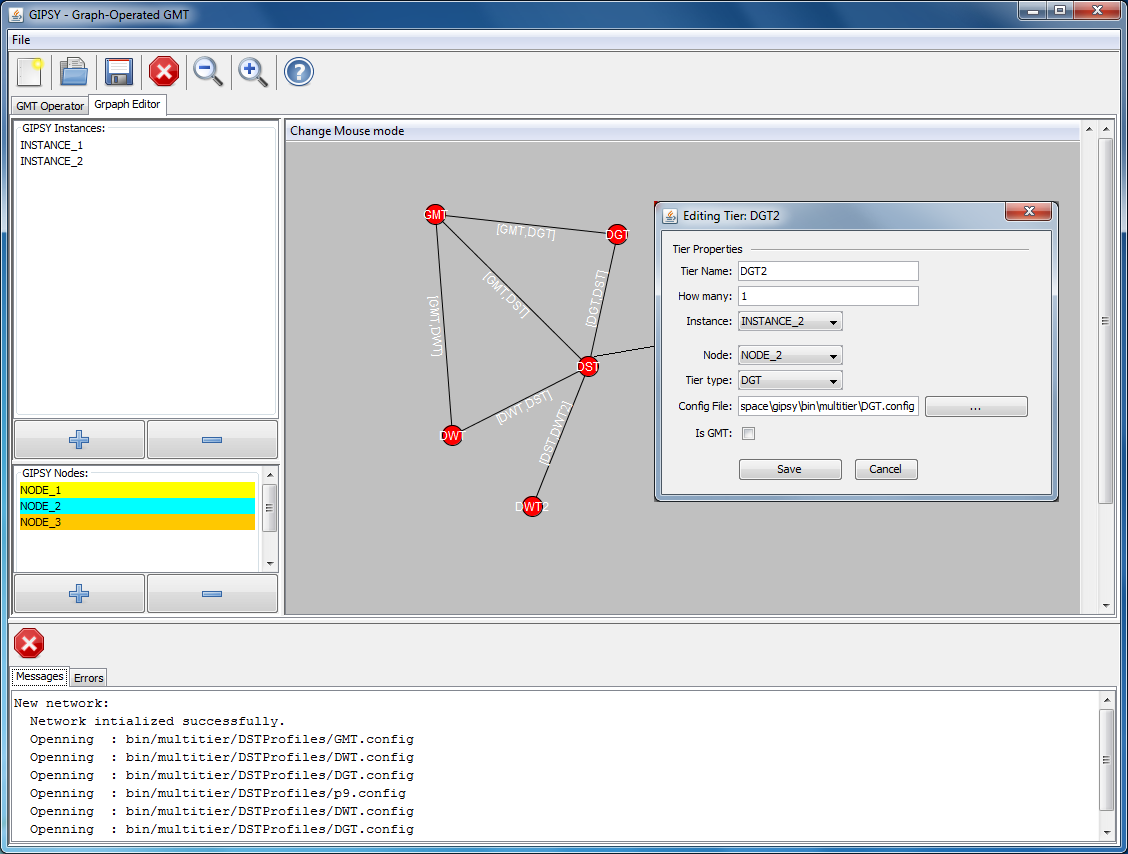} 
			 \label{fig:network-graph-editor}
			}		  
	}
\caption{GMT Operator and Network Graph Editor Views}
\label{fig:operator-net-graph-editor}
\end{figure}

%

In \xf{fig:gmt-visualization-classes} is a set of JUNG-interfaced
classes we produced to integrate with and visually represent,
load, save, and manage {\gipsynetwork}s while in \xf{fig:gmt-data-structure}
we detail the data structures used to internally represent the network
graphs and map them to the {\gipsy} objects and the action items associated
with them. \api{NodeConnection} is a semantically central data structure
that links graph elements representing {\gipsy} tiers (the instances of
\api{GIPSYTier} classes). These connections and tier properties are the
actual representation of the graphs that are saved to and loader from
a name:value paired configuration files (e.g. see \xa{sect:example-stored-graph})
by \api{GraphDataManager}.
A collection of \api{NodeConnection}s is managed by the \api{GraphViewer},
and both \api{NodeConnection} and \api{GIPSYTier} have action items attached to
them that send the aforementioned {\gipsy} commands to actually do the work via the 
visual \api{NodeMenu} and \api{TierMenu}. Every tier has a color and shape
(\api{VertexColorTransformer}, \api{VertexShapeSizeAspect}) attached to it based on the
\api{GIPSYNode} they belong to, so it is easier to differentiate and visualize the computing
resource allocations. When each graph is loaded, mapping is made, and colors determined,
the data structures are handed over to JUNG to do the visual layout.

\begin{figure*}[ht]
\begin{center}
\mbox{
	\subfigure[Visualization Classes]
	{\label{fig:gmt-visualization-classes}
	\includegraphics[width=3in]{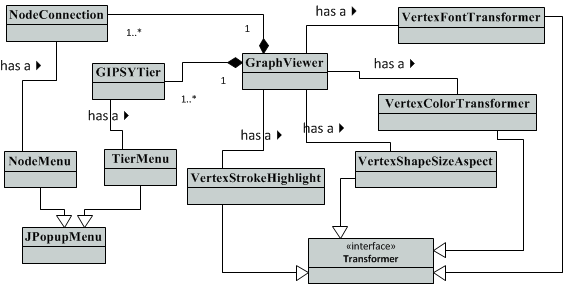}}
	\subfigure[Graph-Related Data Structures]
	{\label{fig:gmt-data-structure}
	\quad
	\includegraphics[width=3in]{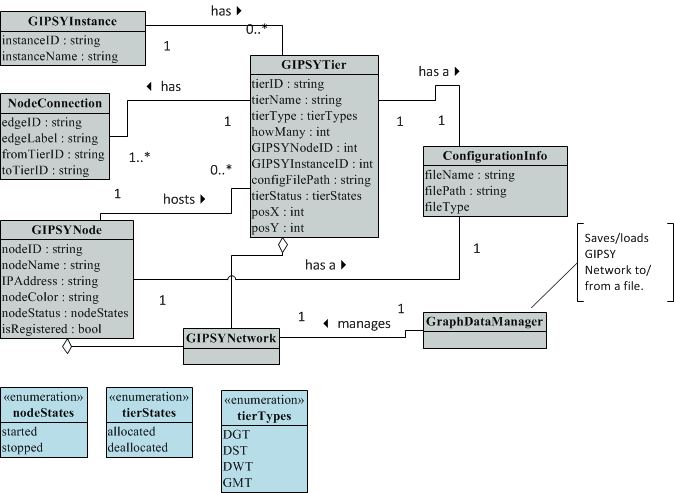}}
}
\caption{UML Class Diagrams of the Graph-Related Design Aspects}
\label{fig:vis-data-structure-class-diags}
\end{center}
\end{figure*}

\subsection{Results and Discussion}
\label{sect:resultsAndDiscussion}

The results are encouraging since they demonstrate the ability of the 
proposed solution to assist in automation of some management functions
{\gipsy} run-time system. We have first tested with the
simulator
\cite{dmf-pdpta07,ji-yi-mcthesis-2011}
which allows to generate different types of demands to 
be computed.
We have then performed some usability testing with another tool, that
was recently adapted to be distributively executed over {\gipsy} -- {\marfcat}.
It is one of the realistic long-running distributed pattern recognition computation
processes test cases (e.g. {\marf}'s pattern recognition pipeline \cite{marf-c3s2e08}
with very large data sets over {\gipsy} for the static code analysis application
for vulnerabilities and weaknesses detection and malware classification 
\cite{marfcat-arxiv,marfcat-sate4-arxiv}. {\marfcat} was made to run completely
over {\gipsy} separating the heavy and light work logic across the generator
and worker tiers. The tool properly starts up all the indicated components, the
network of which were created and the configurations loaded, begins the computation,
logs the output to the console, and while computation proceeds, the tier state is
properly reflected visually.

While us we were the only users of the proposed PoC tool thus far, by
making it public and released along with the {\gipsy} system and via the
demonstration of the tool on the GIPSY simulator and {\marfcat}, we hope to
gather larger feedback on the tool to improve its usability further while
weeding out known bugs.

In \xa{sect:ggmt-demo-manual} we summarize a complete demo procedure/manual
for a creation of a specific application to run over the {\gipsynetwork} for
demonstration purposes.

\paragraph{Platforms Tested.}

We tested the tool in various operating system platforms to ensure the
portability is maintained:
%
\begin{inparaitem}[]
	\item Windows XP SP3 32-bit, Windows 7 32-bit and 64-bit;
	\item Scientific Linux 6.2 32-bit and 64-bit, and Ubuntu Linux 11.03 32-bit under VMware;
	\item MacOS X 10.5 32-bit and 64-bit;
	\item Oracle JDK 6 and 7;
	\item OpenJDK 6;
	\item Apple JDK 6.
\end{inparaitem}

\paragraph{Implications.}

The implications of this work are multifold. First, the usability and management aspects
of the multi-tier {\gipsynetwork} are of obvious mention.
Additionally, having the network
represented and managed as a graph allows for further reasoning and automatic scheduling \cite{ben-hamed-phd-08}
and load-balancing of such a network through the graph analysis.
Thirdly, since {\lucid} is a data-flow language and was shown to have one-to-one
correspondence with the data-flow graphs (DFGs) \cite{paquetThesis,yimin04,flucid-dfg-viz-pst2011},
the tool opens up more possibilities for diagrammatic programming and program-graph-execution-network
translation model for detailed analysis and verification of {\lucid}-based programs with the added
visualization benefit.

\section{Conclusion and Future Work}	
\label{sect:conclusion}

We have presented a graph-based GUI implementation for the simplification of the 
management of the {\gipsy} run-time system components. The presented tool is proving
to be an effective solution assisting with management automation of {\gipsy} 
software artifacts distributed across multiple physical machines 
forming an overlay network. Our solution relies on graph-based programming 
and visualization to represent a {\gipsynetwork}. Each graph node represents 
a {\gipsytier} and is pre-configured and loaded with the information needed 
at run-time. A {\gipsynetwork} can be created, configured and saved to a 
file. The user can establish a connection between pairs of {\gipsytier}s by 
drawing a line to connect two graph nodes. A {\gipsynetwork} can be easily 
bootstrapped and managed on the fly. Many demand generators and workers can 
be allocated as computation is happening. While aiming at increasing the 
usability of the run-time system, our solution allows the user to seamlessly 
inspect the status and properties of {\gipsynode}s and {\gipsytier}s at run-time.

The work presented in this paper is to be extended, thus, additional features 
and improvements are planned. Future work includes a better semantic 
definitions of the graph manipulation actions, so that any operation on a 
graph can be translated more easily into the underlying system's commands
and be verifiable. 
We plan to add observers to any graph element, 
enabling for example to click on a graph link to observe the demands flowing 
across this link at run-time. 
Among the planned future works is the continual extension of the current 
design to support more problem-specific tiers like {\marfcat}, e.g.
genome sequence alignment, and similar computation problems that need a lot
of manual pre-setup to run.
We further plan to allow intra-tool (peer) communication to further allow start up nodes on remote
computers and not only tiers. Additionally, expose an OpenGL-to-Java remote
interface to allow connecting to the tool from any OpenGL-enabled systems remotely,
including mobile devices based on iOS and Android.

%
%
%

\section*{Acknowledgements}

This work in part is supported by NSERC and the Faculty of Engineering and 
Computer Science of Concordia University. We thank reviewers for their 
constructive reviews and feedback.

\bibliographystyle{plain}
\bibliography{graph-based-gmt}

\clearpage
\appendix

\addtocontents{toc}{\cftpagenumbersoff{section}}

\section{Mini Demo User Manual / ``Demo Paper''}
\label{sect:ggmt-demo-manual}

In this section we give a brief overview on how to use the graphical PoC tool to 
create and manage a {\gipsynetwork}. We explain how to create and configure a 
network, how the network is saved/loaded to/from a file, and 
finally, how the network is being managed in the GMT Operator view.

This section is intended for the demonstration of the tool at the conference
with a mini-user manual instructions and operational description.
We will show how to create and start a complete GIPSY network for the specific
{\marfcat}
application and perform a complete run of it. We will release the PoC tools, the
application, and the source code for the audience and community at large
as well.

\subsection{Using the Graph editor}

The graph editor is used to create a new {\gipsynetwork} or to edit an existing one. 

\begin{enumerate}[I.]


\item \textit{Creating/Editing a {\gipsyinstance}}

\xf{fig:creating-gipsy-instance} and \xf{fig:editing-gipsy-instance} show how 
to create a {\gipsyinstance} and edit its information.
By clicking on the add button, the user enters a name for the new {\gipsy} 
instance to create and clicks save.
Double-clicking on an instance in the list of instances allows to edit the 
instance name in an appropriate dialog.


\item \textit{Creating/Editing a GIPSY Node}

To create a new {\gipsynode}, click on the ``Add'' button \xf{fig:add-gipsy-node},
which pops up a dialog to fill in the new node's properties such as the node 
name, IP address and color, see \xf{fig:create-new-gipsy-node}. Upon clicking 
``Save'', the new node will be added to the list as shown in \xf{fig:new-gipsy-node-added}.
To edit an existing node's properties, double-click on an item in the node 
list and a editing dialog will pop up \xf{fig:editing-gipsy-node}.


\begin{figure*}[ht]
\begin{center}
\mbox{
	\subfigure[Creating new GIPSY Instance]
	{\label{fig:creating-gipsy-instance}
	\includegraphics[width=.31\textwidth]{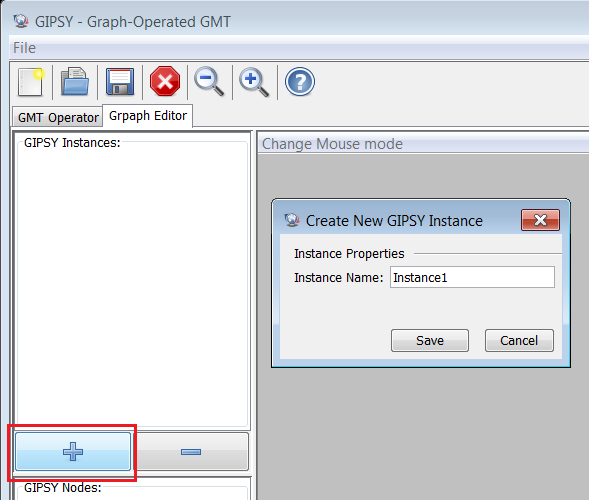}}
	\hspace{5pt}
	\subfigure[Editing GIPSY Instance]
	{\label{fig:editing-gipsy-instance}
	\includegraphics[width=.33\textwidth]{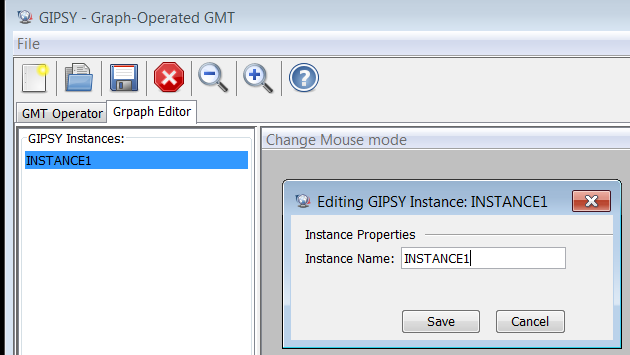}}
	\hspace{5pt}
	\subfigure[Add GIPSY Node]
	{\label{fig:add-gipsy-node}
	\includegraphics[width=.29\textwidth]{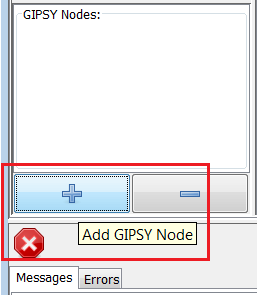}}
}
\caption{Adding a GIPSY Instance and a GIPSY Node}
\label{fig:create-edit-add-gipsy-instance-node}
\end{center}
\end{figure*}

%


\begin{figure*}[ht]
\begin{center}
\mbox{
	\subfigure[Create New GIPSY Node]
	{\label{fig:create-new-gipsy-node}
	\includegraphics[width=.35\textwidth]{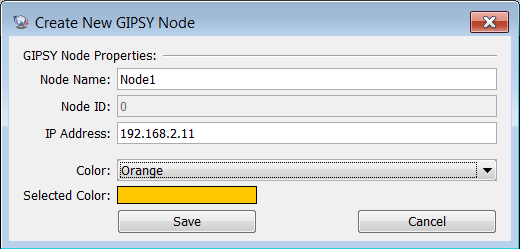}}
	\hspace{2pt}
	\subfigure[New GIPSY Node Added]
	{\label{fig:new-gipsy-node-added}
	\includegraphics[width=.24\textwidth]{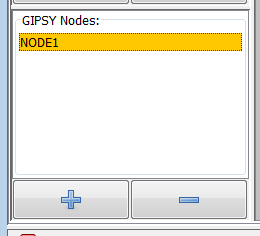}}
	\hspace{2pt}
	\subfigure[Editing GIPSY Node]
	{\label{fig:editing-gipsy-node}
	\includegraphics[width=.35\textwidth]{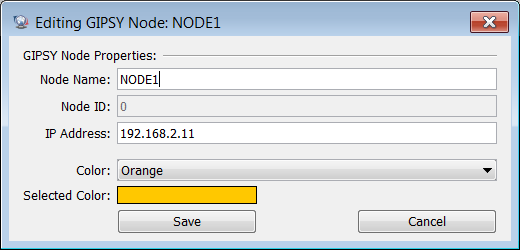}}
}
\caption{Creating and Editing a GIPSY Node}
\label{fig:creating-editing-gipsy-node}
\end{center}
\end{figure*}

\item \textit{Creating/Editing a {\gipsytier}}

While editing, a {\gipsytier} is represented as a red graph node. To create a {\gipsytier}, 
the user must double-click on the highlighted area as shown in
\xf{fig:creating-gipsy-tier}. Then, the tier properties such as the
tier name, how many instances to create, {\gipsyinstance} to which the tier 
belongs to, {\gipsynode} on which the tier will be allocated, and finally a 
configuration file should be specified, cf. \xf{fig:create-configure-gipsy-tier}.
To edit a given tier's properties, right-click on a graph node and select
``Edit Tier Properties'', see \xf{fig:edit-tier-properties}.



\begin{figure*}[ht]
\begin{center}
\mbox{
	\subfigure[Creating a GIPSY Tier]
	{\label{fig:creating-gipsy-tier}
	\includegraphics[width=.49\textwidth]{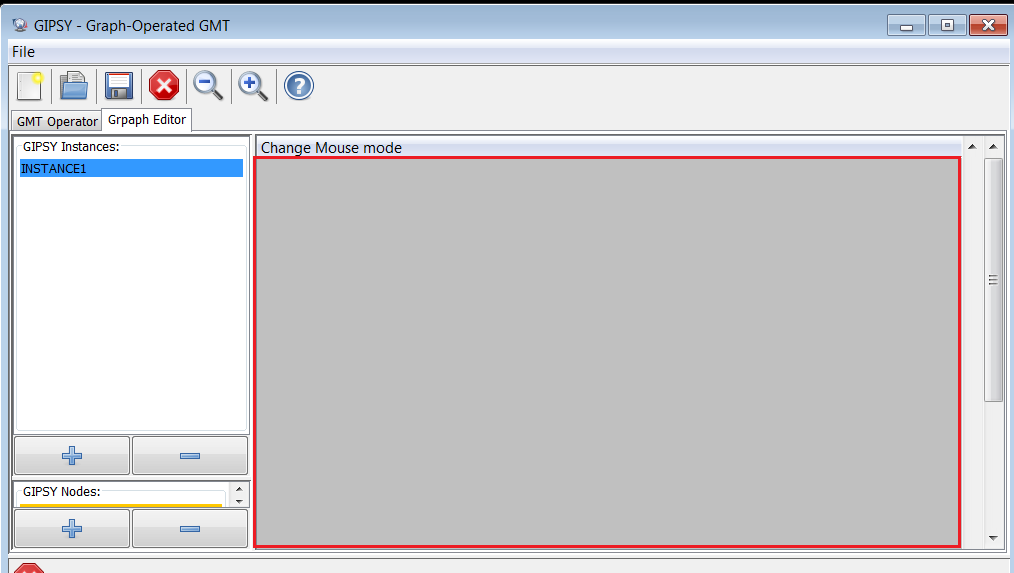}}
	\hspace{2pt}
	\subfigure[Fill In Tier Configuration Properties]
	{\label{fig:configure-tier}
	\includegraphics[width=.48\textwidth]{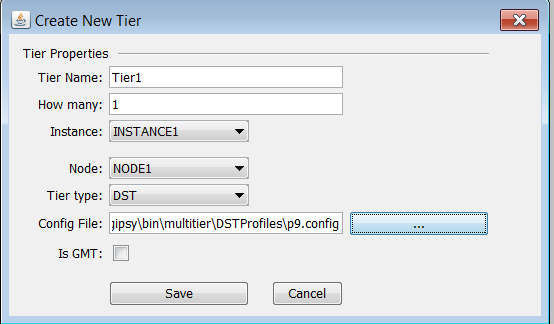}}
}
\caption{Creating and Configuring a GIPSY Tier}
\label{fig:create-configure-gipsy-tier}
\end{center}
\end{figure*}

%

\begin{figure*}[ht]
\begin{center}
\mbox{
	\subfigure[Right Click to Edit Tier Properties]
	{\label{fig:edit-tier-right-click}
	\includegraphics[width=.49\textwidth]{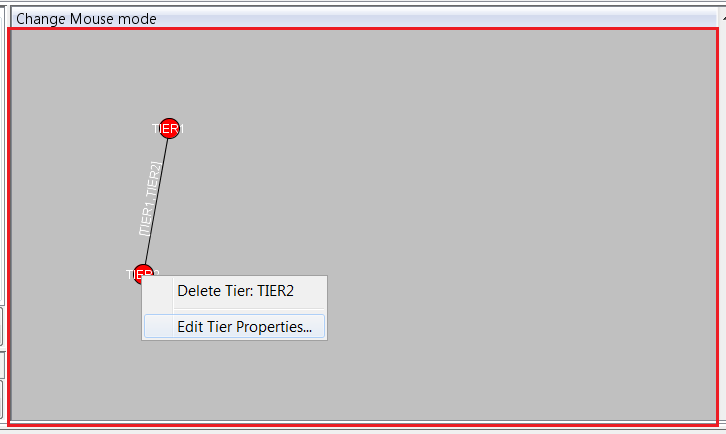}}
	\hspace{2pt}
	\subfigure[Edit Tier Properties]
	{\label{fig:edit-tier-properties}
	\includegraphics[width=.48\textwidth]{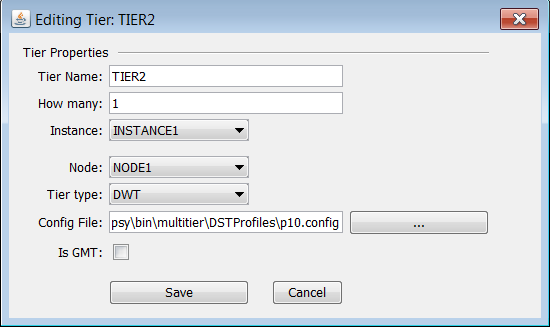}}
}
\caption{Editing Tier Properties}
\label{fig:editing-gipsy-tier-properties}
\end{center}
\end{figure*}

\end{enumerate}

\subsection{Saving/Loading a graph}
\label{sec:SavingLoadingAGraph}

To save/load a network to/form a file, use the save/load buttons located in 
the toolbar, \xf{fig:save-load-graph} and \xf{fig:saving-loading-gipsy-graph}.


\subsection{Using the {\gmt} Operator}
\label{sec:UsingTheGMTOperator}

This feature is implemented in the ``{\gmt} Operator'' tab and enabled
upon loading a valid saved network graph. 
After loading is complete, the graph nodes (GIPSY Tiers) have
the same color as the GIPSY node they belong to. Tiers' shapes,
as mentioned early in this paper, indicate
what GIPSY Tier belongs to what GIPSY Instance.



\begin{figure*}[ht]
\begin{center}
\mbox{
	\subfigure[Saving/Loading Graph]
	{\label{fig:save-load-graph}
	\includegraphics[width=.30\textwidth]{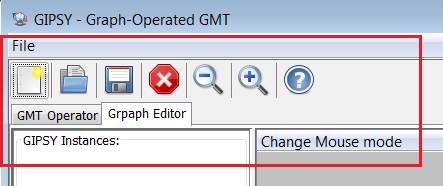}}
	\hspace{2pt}
	\subfigure[Load Graph]
	{\label{fig:load-graph}
	\includegraphics[width=.30\textwidth]{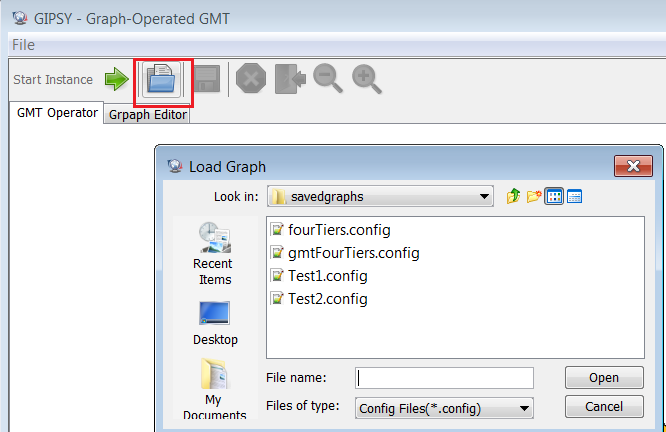}}
	\hspace{2pt}
	\subfigure[Loaded Graph Example]
	{\label{fig:loaded-graph-example}
	\includegraphics[width=.32\textwidth]{images/image029}}
}
\caption{Saving and Loading {\gipsynetwork} Graph}
\label{fig:saving-loading-gipsy-graph}
\end{center}
\end{figure*}

\begin{enumerate}[I.]
\item \textit{Starting/Stopping a {\gipsynode}}
 
To start a {\gipsynode}, right-click on a item in the nodes list and select
``Start Node'', \xf{fig:starting-node}.


After starting the first GIPSY Node, the actions taken
are logged in the log console in ``Messages'' tab.


When the {\gmt} is first started, a new tab is added to the
log console where its activities are logged, see \xf{fig:node-startup-and-registration}.


\begin{figure*}[ht]
\begin{center}
\mbox{
	\subfigure[Starting Node]
	{\label{fig:starting-node}
	\includegraphics[width=.30\textwidth]{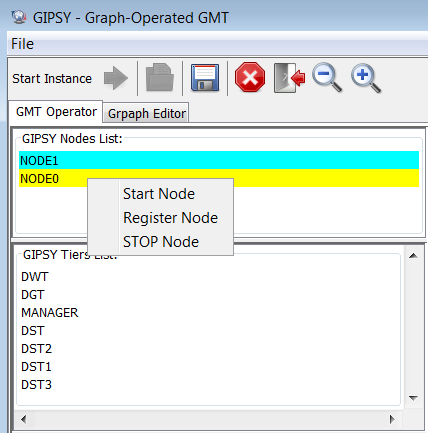}}
	\hspace{2pt}
	\subfigure[Starting Node Messages]
	{\label{fig:starting-node-messages}
	\includegraphics[width=.32\textwidth]{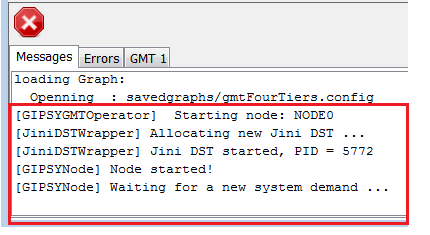}}
	\hspace{2pt}
	\subfigure[Node Registered]
	{\label{fig:node-registration}
	\includegraphics[width=.32\textwidth]{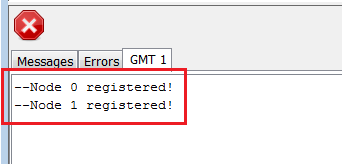}}
}
\caption{Saving and Loading {\gipsynetwork} Graph}
\label{fig:node-startup-and-registration}
\end{center}
\end{figure*}

\item \textit{Allocation/Deallocating {\gipsytier}}

To allocate or deallocate a GIPSY Tier, right-click on a graph
node and select the appropriate action, \xf{fig:start-tier}. The messages and 
action triggered by the allocation/deallocation process are logged and showed 
in the console tab, \xf{fig:start-allocate-console-messages},
\xf{fig:tier-allocation-deallocation}.

\end{enumerate}



\begin{figure*}[ht]
\begin{center}
\mbox{
	\subfigure[Start Tier]
	{\label{fig:start-tier}
	\includegraphics[width=.30\textwidth]{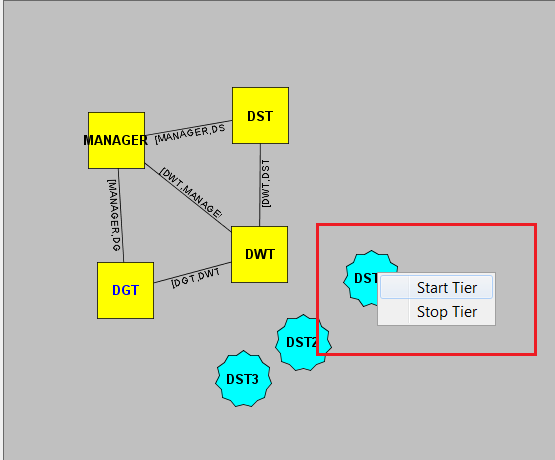}}
	\hspace{2pt}
	\subfigure[Allocation/Startup Console Messages]
	{\label{fig:start-allocate-console-messages}
	\includegraphics[width=.67\textwidth]{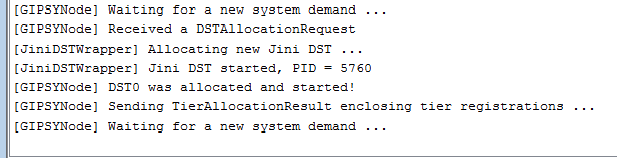}}
}
\caption{Starting a Selected Tier}
\label{fig:tier-selection-and-startup}
\end{center}
\end{figure*}
 
%
%

\begin{figure*}[ht]
\begin{center}
\mbox{
	\subfigure[Allocation of DST]
	{\label{fig:allocate-dst}
	\includegraphics[width=.33\textwidth]{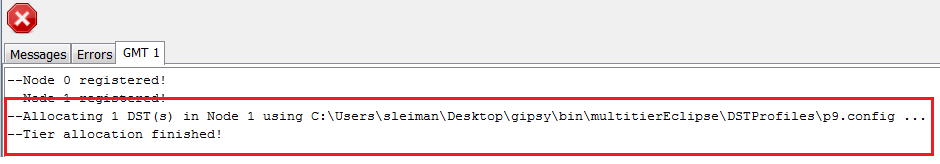}}
	\hspace{2pt}
	\subfigure[Deallocation after a System Demand]
	{\label{fig:deallocate-dst}
	\includegraphics[width=.31\textwidth]{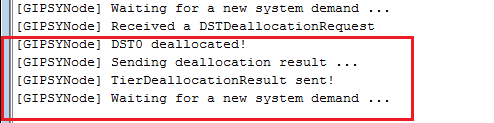}}
	\hspace{2pt}
	\subfigure[Deallocation Completed]
	{\label{fig:deallocation-completed}
	\includegraphics[width=.30\textwidth]{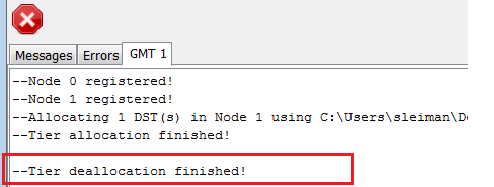}}
}
\caption{Allocation and Deallocation of a DST}
\label{fig:tier-allocation-deallocation}
\end{center}
\end{figure*}

\section{Stored Graph Example}
\label{sect:example-stored-graph}

In the example below is a concrete on-disk representation of the {\gipsynetwork}
graph from \file{marfcat4Some.config} that can be stored and retrieved and executed
instantiating the designed configuration and its connectivity for the {\marfcat}
test case with the corresponding graph in \xf{fig:marfcat4some-plain-net}.

\begin{figure}[htpb]%
	\centering
	\includegraphics[width=.7\columnwidth]{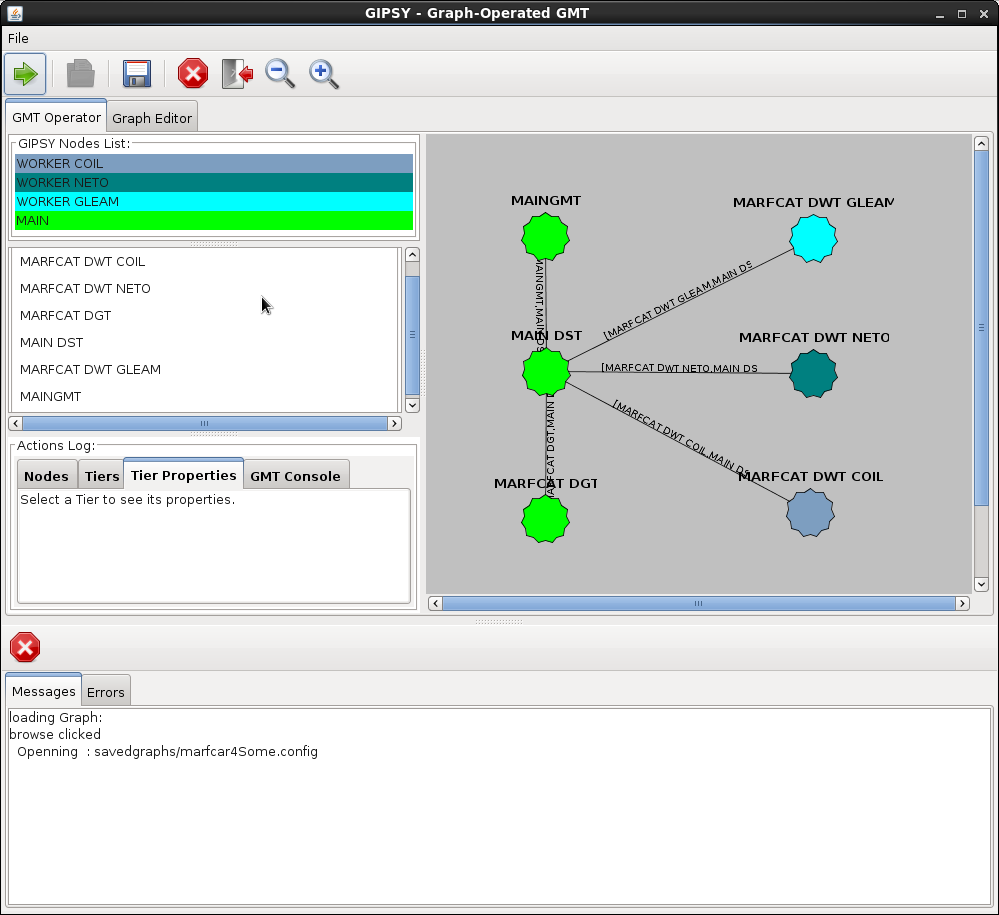}%
	\caption{{\marfcat} {\gipsynetwork} Graph}%
	\label{fig:marfcat4some-plain-net}%
\end{figure}

\tiny
\begin{Verbatim}
#Graph data
#Mon Jun 25 23:04:50 EDT 2012
gipsy.tier.posx.a409cb6b-ad54-41f6-af69-f4ec78628ec6=405 
gipsy.tier.isgmt.89e6605e-5c81-4d95-bf51-8e7576c8dcd7=false 
gipsy.tier.ID.1523569d-25aa-44a4-97d5-370b93790333=1523569d-25aa-44a4-97d5-370b93790333
gipsy.tier.posx.89e6605e-5c81-4d95-bf51-8e7576c8dcd7=409 
gipsy.connection.to.7802414d-437f-4382-9162-15d9d8e735c3=77fd90ec-e49e-4977-a55b-0690e2e0cd08
gipsy.tier.howmany.a409cb6b-ad54-41f6-af69-f4ec78628ec6=1
gipsy.tier.isgmt.c341ab91-161e-419d-a7e4-6206e75ec097=false 
gipsy.tier.howmany.89e6605e-5c81-4d95-bf51-8e7576c8dcd7=1
gipsy.tier.node.1523569d-25aa-44a4-97d5-370b93790333=1
gipsy.tier.posx.c341ab91-161e-419d-a7e4-6206e75ec097=74 
gipsy.tier.howmany.c341ab91-161e-419d-a7e4-6206e75ec097=1
connection_key.0dd7156f-3b40-4df7-b9a9-d1b44eeb4da0=0dd7156f-3b40-4df7-b9a9-d1b44eeb4da0
gipsy.node.name.3=WORKER COIL
gipsy.node.name.2=WORKER NETO
gipsy.tier.posy.1523569d-25aa-44a4-97d5-370b93790333=55 
gipsy.node.name.1=WORKER GLEAM
gipsy.tier.isgmt.77fd90ec-e49e-4977-a55b-0690e2e0cd08=false 
gipsy.node.name.0=MAIN
gipsy.tier.posx.77fd90ec-e49e-4977-a55b-0690e2e0cd08=75 
gipsy.connection.to.0dd7156f-3b40-4df7-b9a9-d1b44eeb4da0=77fd90ec-e49e-4977-a55b-0690e2e0cd08
gipsy.tier.howmany.77fd90ec-e49e-4977-a55b-0690e2e0cd08=1
gipsy.tier.node.29310d89-fd7a-498c-8a47-f80d3b5b9865=0
gipsy.connection.id.1eeea2b5-ec50-4990-a7fe-be48bb5b693e=1eeea2b5-ec50-4990-a7fe-be48bb5b693e
gipsy.tier.tiertype.a409cb6b-ad54-41f6-af69-f4ec78628ec6=DWT
gipsy.tier.posy.29310d89-fd7a-498c-8a47-f80d3b5b9865=53 
gipsy.tier.ID.29310d89-fd7a-498c-8a47-f80d3b5b9865=29310d89-fd7a-498c-8a47-f80d3b5b9865
gipsy.tier.isgmt.1523569d-25aa-44a4-97d5-370b93790333=false 
gipsy.connection.id.6a9e123b-c375-4786-9d15-12faeee54c59=6a9e123b-c375-4786-9d15-12faeee54c59
gipsy.tier.posx.1523569d-25aa-44a4-97d5-370b93790333=409 
gipsy.tier.instance.a409cb6b-ad54-41f6-af69-f4ec78628ec6=8f5d7614-49e9-4c8a-87fe-c8032a5c9009
gipsy.tier.howmany.1523569d-25aa-44a4-97d5-370b93790333=1
gipsy.connection.name.6a9e123b-c375-4786-9d15-12faeee54c59=[MARFCAT DGT,MAIN DST]
gipsy.tier.tiertype.89e6605e-5c81-4d95-bf51-8e7576c8dcd7=DWT
gipsy.tier.tiertype.77fd90ec-e49e-4977-a55b-0690e2e0cd08=DST
gipsy.connection.id.36485580-5acf-4c12-9110-47c60e23ddc9=36485580-5acf-4c12-9110-47c60e23ddc9
gipsy.connection.name.1eeea2b5-ec50-4990-a7fe-be48bb5b693e=[MARFCAT DWT COIL,MAIN DST]
gipsy.tier.tiertype.c341ab91-161e-419d-a7e4-6206e75ec097=DGT
gipsy.tier.isgmt.29310d89-fd7a-498c-8a47-f80d3b5b9865=true 
gipsy.connection.name.36485580-5acf-4c12-9110-47c60e23ddc9=[MARFCAT DWT NETO,MAIN DST]
gipsy.tier.posx.29310d89-fd7a-498c-8a47-f80d3b5b9865=74 
gipsy.tier.instance.89e6605e-5c81-4d95-bf51-8e7576c8dcd7=8f5d7614-49e9-4c8a-87fe-c8032a5c9009
gipsy.connection.from.7802414d-437f-4382-9162-15d9d8e735c3=29310d89-fd7a-498c-8a47-f80d3b5b9865
gipsy.tier.howmany.29310d89-fd7a-498c-8a47-f80d3b5b9865=1
connection_key.6a9e123b-c375-4786-9d15-12faeee54c59=6a9e123b-c375-4786-9d15-12faeee54c59
gipsy.tier.name.a409cb6b-ad54-41f6-af69-f4ec78628ec6=MARFCAT DWT COIL
gipsy.tier.instance.c341ab91-161e-419d-a7e4-6206e75ec097=8f5d7614-49e9-4c8a-87fe-c8032a5c9009
gipsy.tier.tiertype.1523569d-25aa-44a4-97d5-370b93790333=DWT
gipsy.tier.configfile.a409cb6b-ad54-41f6-af69-f4ec78628ec6=bin/multitier/DSTProfiles/marfcatDWT.config
connection_key.1eeea2b5-ec50-4990-a7fe-be48bb5b693e=1eeea2b5-ec50-4990-a7fe-be48bb5b693e
tier_key.a409cb6b-ad54-41f6-af69-f4ec78628ec6=a409cb6b-ad54-41f6-af69-f4ec78628ec6
gipsy.connection.to.6a9e123b-c375-4786-9d15-12faeee54c59=77fd90ec-e49e-4977-a55b-0690e2e0cd08
tier_key.89e6605e-5c81-4d95-bf51-8e7576c8dcd7=89e6605e-5c81-4d95-bf51-8e7576c8dcd7
instance_key.8f5d7614-49e9-4c8a-87fe-c8032a5c9009=8f5d7614-49e9-4c8a-87fe-c8032a5c9009
connection_key.36485580-5acf-4c12-9110-47c60e23ddc9=36485580-5acf-4c12-9110-47c60e23ddc9
gipsy.tier.instance.77fd90ec-e49e-4977-a55b-0690e2e0cd08=8f5d7614-49e9-4c8a-87fe-c8032a5c9009
gipsy.connection.to.1eeea2b5-ec50-4990-a7fe-be48bb5b693e=77fd90ec-e49e-4977-a55b-0690e2e0cd08
gipsy.connection.from.0dd7156f-3b40-4df7-b9a9-d1b44eeb4da0=1523569d-25aa-44a4-97d5-370b93790333
tier_key.c341ab91-161e-419d-a7e4-6206e75ec097=c341ab91-161e-419d-a7e4-6206e75ec097
gipsy.tier.instance.1523569d-25aa-44a4-97d5-370b93790333=8f5d7614-49e9-4c8a-87fe-c8032a5c9009
gipsy.connection.to.36485580-5acf-4c12-9110-47c60e23ddc9=77fd90ec-e49e-4977-a55b-0690e2e0cd08
gipsy.node.ipaddress.3=132.205.8.235
gipsy.node.ipaddress.2=132.205.8.235
gipsy.tier.tiertype.29310d89-fd7a-498c-8a47-f80d3b5b9865=GMT
gipsy.tier.name.89e6605e-5c81-4d95-bf51-8e7576c8dcd7=MARFCAT DWT NETO
gipsy.node.ipaddress.1=132.205.8.235
gipsy.tier.name.77fd90ec-e49e-4977-a55b-0690e2e0cd08=MAIN DST
gipsy.node.ipaddress.0=132.205.8.235
gipsy.tier.configfile.89e6605e-5c81-4d95-bf51-8e7576c8dcd7=bin/multitier/DSTProfiles/marfcatDWT.config
gipsy.tier.configfile.77fd90ec-e49e-4977-a55b-0690e2e0cd08=bin/multitier/DSTProfiles/p9.config
tier_key.77fd90ec-e49e-4977-a55b-0690e2e0cd08=77fd90ec-e49e-4977-a55b-0690e2e0cd08
gipsy.tier.name.c341ab91-161e-419d-a7e4-6206e75ec097=MARFCAT DGT
gipsy.instance.name.8f5d7614-49e9-4c8a-87fe-c8032a5c9009=MARFCAT 4SOME
node_key.3=3
gipsy.tier.configfile.c341ab91-161e-419d-a7e4-6206e75ec097=/local/data/gipsy/cvscheckout/gipsy/gipsy/bin/multitier/marfcatDGT.config
gipsy.node.ID.3=3
node_key.2=2
gipsy.node.ID.2=2
node_key.1=1
gipsy.instance.ID.8f5d7614-49e9-4c8a-87fe-c8032a5c9009=8f5d7614-49e9-4c8a-87fe-c8032a5c9009
node_key.0=0
gipsy.node.ID.1=1
gipsy.node.ID.0=0
gipsy.node.color.3=Lightblue
gipsy.node.color.2=Teal
gipsy.node.color.1=Blue
gipsy.node.color.0=Green
gipsy.tier.name.1523569d-25aa-44a4-97d5-370b93790333=MARFCAT DWT GLEAM
gipsy.tier.configfile.1523569d-25aa-44a4-97d5-370b93790333=/local/data/gipsy/cvscheckout/gipsy/gipsy/bin/multitier/marfcatDWT.config
tier_key.1523569d-25aa-44a4-97d5-370b93790333=1523569d-25aa-44a4-97d5-370b93790333
gipsy.tier.ID.a409cb6b-ad54-41f6-af69-f4ec78628ec6=a409cb6b-ad54-41f6-af69-f4ec78628ec6
gipsy.tier.instance.29310d89-fd7a-498c-8a47-f80d3b5b9865=8f5d7614-49e9-4c8a-87fe-c8032a5c9009
gipsy.tier.node.a409cb6b-ad54-41f6-af69-f4ec78628ec6=3
gipsy.connection.id.7802414d-437f-4382-9162-15d9d8e735c3=7802414d-437f-4382-9162-15d9d8e735c3
gipsy.connection.name.7802414d-437f-4382-9162-15d9d8e735c3=[MAINGMT,MAIN DST]
gipsy.tier.posy.a409cb6b-ad54-41f6-af69-f4ec78628ec6=398 
gipsy.tier.node.c341ab91-161e-419d-a7e4-6206e75ec097=0
tier_key.29310d89-fd7a-498c-8a47-f80d3b5b9865=29310d89-fd7a-498c-8a47-f80d3b5b9865
gipsy.connection.from.6a9e123b-c375-4786-9d15-12faeee54c59=c341ab91-161e-419d-a7e4-6206e75ec097
gipsy.tier.ID.89e6605e-5c81-4d95-bf51-8e7576c8dcd7=89e6605e-5c81-4d95-bf51-8e7576c8dcd7
gipsy.tier.ID.77fd90ec-e49e-4977-a55b-0690e2e0cd08=77fd90ec-e49e-4977-a55b-0690e2e0cd08
gipsy.connection.from.1eeea2b5-ec50-4990-a7fe-be48bb5b693e=a409cb6b-ad54-41f6-af69-f4ec78628ec6
gipsy.tier.posy.c341ab91-161e-419d-a7e4-6206e75ec097=406 
gipsy.connection.id.0dd7156f-3b40-4df7-b9a9-d1b44eeb4da0=0dd7156f-3b40-4df7-b9a9-d1b44eeb4da0
gipsy.tier.ID.c341ab91-161e-419d-a7e4-6206e75ec097=c341ab91-161e-419d-a7e4-6206e75ec097
gipsy.tier.node.89e6605e-5c81-4d95-bf51-8e7576c8dcd7=2
gipsy.tier.node.77fd90ec-e49e-4977-a55b-0690e2e0cd08=0
gipsy.tier.name.29310d89-fd7a-498c-8a47-f80d3b5b9865=MAINGMT
gipsy.connection.from.36485580-5acf-4c12-9110-47c60e23ddc9=89e6605e-5c81-4d95-bf51-8e7576c8dcd7
gipsy.tier.configfile.29310d89-fd7a-498c-8a47-f80d3b5b9865=/local/data/gipsy/cvscheckout/gipsy/gipsy/bin/multitiermarfcat/GMTNode.config
gipsy.connection.name.0dd7156f-3b40-4df7-b9a9-d1b44eeb4da0=[MARFCAT DWT GLEAM,MAIN DST]
connection_key.7802414d-437f-4382-9162-15d9d8e735c3=7802414d-437f-4382-9162-15d9d8e735c3
gipsy.tier.posy.89e6605e-5c81-4d95-bf51-8e7576c8dcd7=224 
gipsy.tier.posy.77fd90ec-e49e-4977-a55b-0690e2e0cd08=222 
gipsy.tier.isgmt.a409cb6b-ad54-41f6-af69-f4ec78628ec6=false 
\end{Verbatim}
\normalsize

\end{document}

%% file: commands.tex
\newcommand{\xf}[1]{Figure~\ref{#1}}

\newcommand{\xs}[1]{Section~\ref{#1}}
\newcommand{\xa}[1]{Appendix~\ref{#1}}

\newcommand{\xl}[1]{Listing~\ref{#1}}

\newcommand{\gipsy}{{GIPSY\index{GIPSY}}}

\newcommand{\gmt}{{GMT\index{GMT}}}
\newcommand{\dst}{{DST\index{DST}}}
\newcommand{\dgt}{{DGT\index{DGT}}}
\newcommand{\dwt}{{DWT\index{DWT}}}

\newcommand{\gipsynode}{{GIPSY node\index{GIPSY Node}\index{Frameworks!GIPSY Node}}}
\newcommand{\gipsynetwork}{{GIPSY network\index{GIPSY Network}\index{Frameworks!GIPSY Network}}}
\newcommand{\gipsyinstance}{{GIPSY instance\index{GIPSY Instance}\index{Frameworks!GIPSY Instance}}}
\newcommand{\gipsytier}{{GIPSY tier\index{GIPSY Tier}\index{Frameworks!GIPSY Tier}}}

\newcommand{\lucid}{{Lucid\index{Lucid}}}

\newcommand{\file}[1]{\url{#1}\index{Files!#1}}

\newcommand{\api}[1]{\texttt{#1}\index{API!#1}}

\newcommand{\marf}[0]{MARF\index{MARF}\index{Frameworks!MARF}\index{Libraries!MARF}}

\newcommand{\marfcat}{MARFCAT\index{MARFCAT}\index{MARF!Applications!MARFCAT}}

\newcommand{\lucidL}[1]{{$\mathit{Lucid}$}($L$) }

		{}

\def\myvert{\raise 2.27pt \hbox{\vrule depth 0pt height 8pt width 0.2mm}}
\def\myarrow{\hspace*{0.43mm}%
             \raise 2.29pt\hbox{\vrule depth 0pt height 8pt width 0.16mm}%
             \hspace*{-0.32mm}%
             $\longrightarrow$
             \ %
             }


%% file: graph-based-gmt.bbl
\begin{thebibliography}{10}

\bibitem{logical-reasoning-with-diagrams}
Gerard Allwein and Jon Barwise, editors.
\newblock {\em Logical reasoning with diagrams}.
\newblock Oxford University Press, Inc., New York, NY, USA, 1996.

\bibitem{graph-transform-visual-languages}
R.~Bardohl, M.~Minas, G.~Taentzer, and A.~Sch\"{u}rr.
\newblock Application of graph transformation to visual languages.
\newblock In {\em Handbook of Graph Grammars and Computing by Graph
  Transformation: Applications, Languages, and Tools}, volume~2, pages
  105--180. World Scientific Publishing Co., Inc., River Edge, NJ, USA, 1999.

\bibitem{yimin04}
Yimin Ding.
\newblock Automated translation between graphical and textual representations
  of intensional programs in the {GIPSY}.
\newblock Master's thesis, Department of Computer Science and Software
  Engineering, Concordia University, Montreal, Canada, June 2004.
\newblock
  \url{http://newton.cs.concordia.ca/~paquet/filetransfer/publications/theses/DingYiminMSc2004.pdf}.

\bibitem{head-first-patterns-2004}
Eric Freeman, Elisabeth Freeman, Kathy Sierra, and Bert Bates.
\newblock {\em Head First Design Patterns}.
\newblock O'Reilly, first edition, October 2004.
\newblock \url{http://www.oreilly.com/catalog/hfdesignpat/toc.pdf},
  \url{http://www.oreilly.com/catalog/hfdesignpat/chapter/index.html}.

\bibitem{java-reflection}
Dale Green.
\newblock Trail: Java reflection {API}.
\newblock [online], 2001--2012.
\newblock \url{http://docs.oracle.com/javase/tutorial/reflect/index.html}.

\bibitem{ben-hamed-phd-08}
Khaled M.~Ben Hamed.
\newblock {\em Multidimensional Programs on Distributed Parallel Computers:
  Analysis and Implementation}.
\newblock PhD thesis, Computer Science, the University of New Brunswick,
  February 2008.

\bibitem{ji-yi-mcthesis-2011}
Yi~Ji.
\newblock Scalability evaluation of the {GIPSY} runtime system.
\newblock Master's thesis, Department of Computer Science and Software
  Engineering, Concordia University, Montreal, Canada, March 2011.

\bibitem{unifying-refactoring-jini-jms-dms}
Yi~Ji, Serguei~A. Mokhov, and Joey Paquet.
\newblock Unifying and refactoring {DMF} to support concurrent {Jini} and {JMS}
  {DMS} in {GIPSY}.
\newblock In Bipin~C. Desai, Sudhir~P. Mudur, and Emil~I. Vassev, editors, {\em
  Proceedings of the Fifth International C* Conference on Computer Science and
  Software Engineering (C3S2E'12)}, pages 36--44, New York, NY, USA, June
  2010--2012. ACM.
\newblock Online e-print \url{http://arxiv.org/abs/1012.2860}.

\bibitem{jung-url}
{JUNG Project}.
\newblock {Java Universal Network/Graph Framework}.
\newblock [online], 2003--2012.
\newblock \url{http://jung.sourceforge.net/}, last viewed January 2012.

\bibitem{diagramatic-formal-system-euclidean}
N.~G. Miller.
\newblock {\em A Diagrammatic Formal System for Euclidean Geometry}.
\newblock PhD thesis, Cornell University, U.S.A, 2001.

\bibitem{marf-c3s2e08}
Serguei~A. Mokhov.
\newblock Study of best algorithm combinations for speech processing tasks in
  machine learning using median vs. mean clusters in {MARF}.
\newblock In Bipin~C. Desai, editor, {\em Proceedings of C3S2E'08}, pages
  29--43, Montreal, Quebec, Canada, May 2008. ACM.

\bibitem{marfcat-arxiv}
Serguei~A. Mokhov.
\newblock The use of machine learning with signal- and {NLP} processing of
  source code to fingerprint, detect, and classify vulnerabilities and
  weaknesses with {MARFCAT}.
\newblock [online], October 2010.
\newblock Online at \url{http://arxiv.org/abs/1010.2511}.

\bibitem{flucid-dfg-viz-pst2011}
Serguei~A. Mokhov, Joey Paquet, and Mourad Debbabi.
\newblock On the need for data flow graph visualization of {Forensic Lucid}
  programs and forensic evidence, and their evaluation by {GIPSY}.
\newblock In {\em Proceedings of the Ninth Annual International Conference on
  Privacy, Security and Trust (PST), 2011}, pages 120--123. IEEE Computer
  Society, July 2011.
\newblock Short paper; full version online at
  \url{http://arxiv.org/abs/1009.5423}.

\bibitem{marfcat-sate4-arxiv}
Serguei~A. Mokhov, Joey Paquet, Mourad Debbabi, and Yankui Sun.
\newblock {MARFCAT}: Transitioning to binary and larger data sets of {SATE IV}.
\newblock [online], May 2012.
\newblock Being finalized for NIST publication; online at
  \url{http://arxiv.org/abs/1207.3718}.

\bibitem{o-madadhain-fisher-2003}
J.~O'Madadhain et~al.
\newblock The {JUNG (Java Universal Network/Graph)} framework.
\newblock Technical Report UCIICS-03-17, School of Information and Computer
  Science, University of California, Irvine, 2003.

\bibitem{paquetThesis}
Joey Paquet.
\newblock {\em Scientific Intensional Programming}.
\newblock PhD thesis, Department of Computer Science, Laval University,
  Sainte-Foy, Canada, 1999.

\bibitem{gipsy-multi-tier-secasa09}
Joey Paquet.
\newblock Distributed eductive execution of hybrid intensional programs.
\newblock In {\em Proceedings of the 33rd Annual IEEE International Computer
  Software and Applications Conference ({COMPSAC}'09)}, pages 218--224,
  Seattle, Washington, USA, July 2009. IEEE Computer Society.

\bibitem{dmf-pdpta07}
Amir~Hossein Pourteymour, Emil Vassev, and Joey Paquet.
\newblock Towards a new demand-driven message-oriented middleware in {GIPSY}.
\newblock In {\em Proceedings of {PDPTA 2007}}, pages 91--97, Las Vegas, USA,
  June 2007. {PDPTA}, CSREA Press.

\bibitem{visual-config-representation}
Phan~C. Vinh and Jonathan~P. Bowen.
\newblock On the visual representation of configuration in reconfigurable
  computing.
\newblock {\em Electron. Notes Theor. Comput. Sci.}, 109:3--15, 2004.

\bibitem{sim-viz-resource-alloc-control}
Chunfang Zheng and J.~Robert Heath.
\newblock Simulation and visualization of resource allocation, control, and
  load balancing procedures for a multiprocessor architecture.
\newblock In {\em MS'06: Proceedings of the 17th IASTED international
  conference on Modelling and simulation}, pages 382--387, Anaheim, CA, USA,
  2006. ACTA Press.

\end{thebibliography}
